\documentclass[prd,twocolumn,showpacs,nofootinbib,floatfix]{revtex4}
\usepackage{amsmath}

\begin{document}


\title{Using Full Information When Computing Modes of Post-Newtonian
  Waveforms From Inspiralling Compact Binaries in Circular Orbit}


\date{\today}


\author{Lawrence E. Kidder}


\affiliation{Center for Radiophysics and Space Research, Cornell
  University, Ithaca, New York, 14853}


\email{kidder@astro.cornell.edu}


\begin{abstract}
The increasing sophistication and accuracy of numerical simulations of
compact binaries (especially binary black holes) presents the
opportunity to test the regime in which post-Newtonian (PN)
predictions for the emitted gravitational waves are accurate.  In
order to confront numerical results with those of post-Newtonian
theory, it is convenient to compare multipolar decompositions of the
two waveforms.  It is pointed out here that the individual modes can
be computed to higher post-Newtonian order by examining the radiative
multipole moments of the system, rather than by decomposing the 2.5PN
polarization waveforms.  In particular, the dominant ($l = 2, m = \pm
2$) mode {\em can be computed to 3PN order}.  Individual modes are
computed to as high a post-Newtonian order as possible given previous
post-Newtonian results.
\end{abstract}




\maketitle


\section{Introduction}

The first generation of laser interferometric gravitational wave
detectors is now operating at or near their design
sensitivities~\cite{Barish:1999, Waldman:2006, Acernese:2002,
  Acernese-etal:2006, Hild:2006}. One of the most promising sources
that they may detect are the inspiral and merger of compact binaries
systems with black holes or neutron stars.  One of the primary means
of detecting and interpreting the signals from such systems will be
the use of a matched filtering technique, which requires accurate
templates with which to match a theoretical model to a noisy
signal. Until the last several orbits prior to merger, the
post-Newtonian (PN) approximation is expected to be sufficient to
generate accurate templates, while numerical relativity will be needed
to help construct accurate templates covering the late inspiral and
merger phases.

Estimates of the accuracy of post-Newtonian templates and the effect
of this accuracy on detecting gravitational waves have led the
post-Newtonian expansion to be pushed to high
order~\cite{cutler_etal93, Tagoshi94, Krolak95, Damour98}.  Currently
the equations of motion and the gravitational wave energy flux are
known through 3.5PN order~\cite{Blanchet02a, Blanchet04a}, and the
polarization waveform amplitudes are known through 2.5PN
order~\cite{Arun2004,Kidder07}.  As the post-Newtonian expansion is a
slow-motion, weak-field approximation to general relativity, and the
motion is not so slow and the fields not so weak during the last
orbits prior to merger, it is legitimate to ask where the
post-Newtonian expansion breaks down.

Recent breakthroughs in numerical relativity~\cite{Pretorius2005a,
Pretorius2006, Campanelli2006a, Baker2006a} have finally provided the
possibility of testing the validity of the post-Newtonian expansion in
the late inspiral.  Initial studies focused on qualitative comparisons
between post-Newtonian generated waveforms and numerical simulations
of the final two to four orbits prior to
merger~\cite{Buonanno-Cook-Pretorius:2007, Berti2007, Schnittman2007}.
More recently, attempts have been made to quantify the difference
between post-Newtonian generated waveforms and numerical simulations
of a non-spinning equal-mass binary inspiral lasting more than eight
orbits~\cite{Baker2006d, Hannam2007, Boyle2007}.  As long inspiral
simulations are computationally expensive to perform, it is important
to quantify the errors in the post-Newtonian generated waveforms and
determine how they influence the detection of gravitational waves.
One would also like to know the accuracy with which the parameters
such as the individual masses and spins of the binary can be
determined from the observed waveform.

One of the principal ways of quantifying the accuracy of the
post-Newtonian approximation is to compare the amplitude and phase of
the gravitational waveform with that from numerical simulations.
Unlike gravitational wave detection which is sensitive only to the
waveform in one particular direction, comparisons between numerical
simulations and post-Newtonian waveforms can be performed over the
entire sphere.  Therefore it is convenient to project the waveforms
onto spin-weighted spherical harmonics and compare the individual
components.  Previous studies~\cite{Buonanno-Cook-Pretorius:2007,
Berti2007}, however, have computed the spin-weighted spherical
harmonic components of the post-Newtonian amplitude in such a way
that has not made full use of current post-Newtonian results.

This situation has arisen primarily because of how the post-Newtonian
gravitational wave amplitude is presented.  Often the final result of
the post-Newtonian computation of the waveform is given as the
polarization waveforms as a function of the direction to the observer
(as this is most useful when constructing templates).  Currently the
polarization waveforms are known to 2.5PN order in
amplitude~\cite{Arun2004, Kidder07}.  In previous comparisons between
post-Newtonian and numerical waveforms, it was the polarization
waveforms that were projected into spin-weighted spherical harmonics
in order to determine the post-Newtonian components.  The
post-Newtonian polarization waveforms are computed by truncating an
infinite sum of radiative multipole moments at a given post-Newtonian
order~\cite{thorne80}.  As will be demonstrated in this paper, it
makes more sense to project the waveform onto spin-weighted spherical
harmonics {\em prior} to truncating the series as this retains the
full information currently known.  In particular, the dominant
$(l=2,m=2)$ component of the waveform {\em can be computed to 3PN
order}.

While knowledge of a particular spin-weighted components of a
gravitational waveform is of limited value for computing templates for
detection and characterization of gravitational waves, it can be used
to quantify the difference between post-Newtonian and numerical
waveforms.  In addition, if the 3PN correction to the amplitude of the
$(l=2,m=2)$ component significantly improves the agreement between the
post-Newtonian and numerical waveforms, this would suggest it might be
worth the effort needed to compute the polarization waveforms to 3PN
order.  It has been shown that higher post-Newtonian corrections in
the amplitude can improve detection rates~\cite{VanDenBroeck06,
VanDenBroeck07a, Arun07a} as well as improve parameter
extraction~\cite{Sintes00a, Sintes00b, Moore02, Hellings03,
VanDenBroeck07b, Arun07b, Trias07}.

The relationship between the gravitational waveform and radiative
multipole moments is reviewed in Sec.~\ref{sec:grav-wavef-radi}.  It
is pointed out that the the spin-weighted spherical harmonic
components of the waveform can be computed directly from the radiative
mulitpole moments of the system in Sec.~\ref{sec:polar-wavef-modes}.
Current post-Newtonian results that affect the computation of the
radiative multipole moments are summarized in
Sec.~\ref{sec:summary-current-post}.  The radiative multipole moments
are computed to as high a post-Newtonian order as possible in
Sec.~\ref{sec:results}.  Conclusions are presented in
Sec.~\ref{sec:conclusions}.


\section{Gravitational Waveform and Spin-weighted spherical harmonics}
\label{sec:grav-wavef-radi}

\subsection{Notation}

Let $X^\mu = (cT,X,Y,Z)$ be coordinates in an asymptotically flat
radiative coordinate system, with ${\vec e_T}$ the unit timelike
normal, and $({\vec e_X},{\vec e_Y},{\vec e_Z})$ the spatial
orthonormal coordinate basis vectors.  Let $(cT,R,\Theta,\Phi)$ be the
corresponding spherical coordinate system with corresponding basis
vectors $({\vec e_R},{\vec e_\Theta},{\vec e_\Phi})$. Let $T_R = T -
R/c$ denote retarded time in radiative coordinates.  

Greek letters are used for spacetime indices, Latin letters for
spatial indices.  As indices can be raised or lowered using the
Minkowski metric, components of tensors will typically have down
indices.  Let $N_i$ be a component of the unit radial vector ${\vec
e_R}$.  A capital letter for an index denotes a multi-index (i.e. $T_L
= T_{i_1 i_2 \cdots i_\ell}$). For a vector $V$, let $V_L$ be a
product of components of the vector (i.e. $V_L = V_{i_1} V_{i_2}
\cdots V_{i_\ell}$).  Repeated spatial indices (including
multi-indices) are to be summed over.  Tensors that are fully
symmetric and trace-free (STF) will be denoted with capital script
letters (e.g. ${\cal I}_{ij}$).  Symmetrization, anti-symmetrization,
and STF projection of indices are denoted by $()$, $[]$, and $<>$
respectively, with underlined indices between the delimiters to be
excluded from the operation (e.g. $2 T_{(a\underline{b}c)} = T_{abc} +
T_{cba}$).
  
Let $({\vec k},{\vec l},{\vec m})$ be a set of null vectors defined
by:
\begin{eqnarray}
{\vec k} &=& \frac{1}{\sqrt{2}} \left( {\vec e_T} + {\vec e_R} \right), \\
{\vec l} &=& \frac{1}{\sqrt{2}} \left( {\vec e_T} - {\vec e_R} \right), \\
{\vec m} &=& \frac{1}{\sqrt{2}} \left( {\vec e_\Theta} + i {\vec e_\Phi}
 \right). 
\end{eqnarray}

The spin-weighted spherical harmonics are defined in terms of the
Wigner $d$-functions by
\begin{equation}
 {}_{-s}Y^{\ell m}(\Theta,\Phi) = (-1)^s \sqrt{\frac{2 \ell + 1}{4 \pi}}
d^\ell_{ms}(\Theta) e^{i m \Phi},
\end{equation}
where
\begin{eqnarray}
 d^\ell_{ms}(\Theta)& &= \sqrt{(\ell + m)! (\ell - m)! (\ell + s)! (\ell - s)!}
 \nonumber \\*  \times
\sum_{k = k_i}^{k_f}  & &
 \frac{(-1)^k (\sin{\frac{\Theta}{2}})^{2 k + s - m}
(\cos{\frac{\Theta}{2}})^{2 \ell + m -s - 2 k}}{ k! (\ell + m - k)! 
(\ell - s - k)! (s - m + k)!},
\end{eqnarray}
where $k_i$ = max$(0,m-s)$ and $k_f$ = min$(\ell + m,\ell -s)$.


\subsection{Polarization Waveforms and Modes}
\label{sec:polar-wavef-modes}

The asymptotic waveform $h^{TT}_{ij}$ can be
decomposed into two sets of symmetric trace-free (STF) radiative
multipole moments as~\cite{thorne80}
\begin{eqnarray}
\label{eq:hFromSTF}
h^{TT}_{ij} &=& \frac{4G}{c^2 R} \Pi_{ijmn} \sum_{\ell = 2}^\infty
\left\{ \frac{1}{c^\ell \ell !} {\cal U}_{mn L-2}(T_R) N_{L-2} \right.
\nonumber \\* && + \left.
\frac{2 \ell}{c^{\ell + 1} (\ell + 1)!} \epsilon_{pq(m}
{\cal V}_{n)pL-2}(T_R)  N_{qL-2} \right\}.
\end{eqnarray}
Here ${\cal U}_L(T_R)$ are the mass-type moments and ${\cal V}_L(T_R)$
are the current-type moments.  In Sec.~\ref{sec:post-newt-comp} these
radiative multipole moments will be related to multipole moments
describing the source in the near-zone.  Note that higher multipole
moments contribute to the waveform at higher post-Newtonian order, so
that at any finite post-Newtonian order, only a finite number of
multipoles contribute.  The transverse-traceless (TT) projection
operator $\Pi_{ijmn}$ is given by
\begin{equation}
\Pi_{ijmn} = P_{im} P_{jn} - \frac{1}{2} P_{ij} P_{mn},
\end{equation}
where $P_{ij} = \delta_{ij} - N_i N_j$.

Given an orthonormal triad $({\vec N},{\vec P},{\vec Q})$, the
polarization waveforms can be given by
\begin{eqnarray}
h_+ &=& \frac{1}{2} \left( P_m P_n - Q_m Q_n \right) h^{TT}_{mn}, \\
h_\times &=& \frac{1}{2} \left( P_m Q_n + P_n Q_m \right) h^{TT}_{mn}.
\end{eqnarray}
A natural (but by no means unique\footnote{The 2.5PN polarization
waveforms in~\cite{Arun2004} are defined with ${\vec P} = -{\vec
e_\Phi}$ and ${\vec Q} = {\vec e_\Theta}$ evaluated at $\Theta = i$,
$\Phi=\pi/2$.}) choice for the triad is ${\vec P} = {\vec e_\Theta}$
and ${\vec Q} = {\vec e_\Phi}$.  It is then straightforward to show
that
\begin{equation}
\label{eq:ComplexPolarization}
h_+ - i h_\times = m^\ast_m m^\ast_n h^{TT}_{mn},
\end{equation}
where $\ast$ denotes complex conjugation.  It will now be shown how
$h_+ - i h_\times$ can be decomposed into modes using spin-weighted
spherical harmonics of weight -2
\begin{equation}
\label{eq:Decomposition}
h_+ - i h_\times = \sum_{\ell = 2}^\infty \sum_{m = - \ell}^\ell h^{\ell m} 
{}_{-2}Y^{\ell m} \left( \Theta, \Phi \right).
\end{equation}

An alternative expression for the waveform is given by~\cite{thorne80}
\begin{eqnarray}
\label{eq:hFromCoefficients}
h^{TT}_{ij} &=& \frac{G}{c^2 R} \sum_{\ell = 2}^\infty \sum_{m = - \ell}^\ell
\left\{ \frac{1}{c^\ell} U^{\ell m}(T_R) T^{E2,\ell m}_{ij} \right.
\nonumber \\* && + \left. 
\frac{1}{c^{\ell + 1}} V^{\ell m}(T_R) T^{B2,\ell m}_{ij} \right\} ,
\end{eqnarray}
where $T^{E2,\ell m}_{ij}$ and $T^{B2,\ell m}_{ij}$ are pure-spin
tensor harmonics, and where the mass multipole moments $U^{\ell
  m}(T_R)$ and current multipole moments $V^{\ell m}(T_R)$ are related
to their STF counterparts by~\cite{thorne80}
\begin{eqnarray}
U^{\ell m} &=& \frac{16 \pi}{(2 \ell + 1)!!} \sqrt{\frac{(\ell + 1)(\ell + 2)}
{2 \ell (\ell - 1)}} {\cal U}_L {\cal Y}_L^{\ell m \ast},\\
V^{\ell m} &=& \frac{-32 \pi \ell}{(2 \ell + 1)!!} 
\sqrt{\frac{(\ell + 2)}
{2 \ell (\ell + 1)(\ell - 1)}} {\cal V}_L {\cal Y}_L^{\ell m \ast},
\end{eqnarray}
where ${\cal Y}_L^{\ell m}$ are the STF spherical harmonics which are
related to the scalar spherical harmonics by
\begin{equation}
Y^{\ell m}(\Theta,\Phi) = {\cal Y}_L^{\ell m} N_L .
\end{equation}
The pure-spin tensor harmonics are related to the spin-weighted
spherical harmonics by~\cite{thorne80}
\begin{eqnarray}
\label{eq:pureSpinToSpinWeighted}
T^{E2,\ell m}_{ij} &=& \frac{1}{\sqrt{2}} \left( {}_{-2}Y^{\ell m} m_i m_j +  
{}_{2}Y^{\ell m} m^\ast_i m^\ast_j \right), \\
T^{B2,\ell m}_{ij} &=&  \frac{-i}{\sqrt{2}} \left( {}_{-2}Y^{\ell m} m_i m_j 
- {}_{2}Y^{\ell m} m^\ast_i m^\ast_j \right).
\end{eqnarray}
Combining Eqs.~(\ref{eq:ComplexPolarization}), (\ref{eq:hFromCoefficients}),
and (\ref{eq:pureSpinToSpinWeighted}) yields
\begin{eqnarray}
h_+ - i h_\times  &=& 
\frac{G}{c^2 R} \sum_{\ell = 2}^\infty \sum_{m = - \ell}^\ell
\left\{ \frac{1}{c^\ell \sqrt{2}} U^{\ell m}(T_R) \right. 
\nonumber \\* && - \left. 
\frac{i}{c^{\ell + 1} \sqrt{2}} V^{\ell m}(T_R)  \right\} 
{}_{-2}Y^{\ell m} \left( \Theta, \Phi \right).
\end{eqnarray}
Thus the spin-weighted spherical harmonic components of the waveform are
given by
\begin{equation}
\label{eq:ComponentsFromRadiativeMultipoles}
 h^{\ell m} = \frac{G}{\sqrt{2} R c^{\ell + 2} } \left( U^{\ell m}(T_R) 
 - \frac{i}{c} V^{\ell m}(T_R) \right).
\end{equation}

Note that the spin-weighted spherical harmonic components can be
computed {\em directly} from the radiative multipole moments.  It is
not necessary to compute the waveform as a function of position, and
then project out the components as is commonly done in the recent
literature.  In fact, not only are these extra steps unnecessary, they
cause individual components to be truncated at a lower post-Newtonian
order than they need to be.  In particular, the dominant $h^{22}$
component is truncated to 2.5PN order, when it can be computed (as
will be shown below) to 3PN order directly from the radiative mass
quadrupole moment.  

This situation arises as the complete polarization waveforms $h_+$ and
$h_\times$ are known only through 2.5PN order.  But as discussed
in~\cite{Arun2004}, the computation of the 3PN waveform is not limited
by the post-Newtonian order of the dominant mass quadrupole
contribution, but by the order of higher multipoles.  Therefore to
obtain the spin-weighted spherical harmonic components of the waveform
to as high a post-Newtonian order as possible, they should be computed
directly from the radiative multipole moments.  As discussed in
Sec.~\ref{sec:conclusions} below, this will allow post-Newtonian
results to be probed to a higher order when comparing with numerical
simulations.  Table~\ref{tab:pnOrder} summarizes the post-Newtonian
order to which each radiative multipole moment needs to computed to
obtain the full 2.5PN order polarization waveforms, and the
post-Newtonian order to which they are currently known.

\begin{table}
\caption{\label{tab:pnOrder}The post-Newtonian order to which each
  radiative multipole moment must be computed to obtain the full 2.5PN
  polarization waveforms $h_+$ and $h_\times$, and the order to which
  they are known.}
\begin{ruledtabular}
\begin{tabular}{ccc}
Radiative Multipole & PN order for $h_{+,\times}$ & Known PN order \\
${\cal I}_{ij}$ & 2.5 & 3 \\
${\cal J}_{ij}$ & 2 & 2 \\
${\cal I}_{ijk}$ & 2 & 2 \\
${\cal J}_{ijk}$ & 1.5 & 1.5
\footnote{The radiative current octupole could be computed to 2PN order by a 
computation of the 2PN correction to the source current octupole.} \\
${\cal I}_{ijkl}$ & 1.5 & 2 \\
${\cal J}_{ijkl}$ & 1 & 1 \\
${\cal I}_{ijklm}$ & 1 & 1 \\
${\cal J}_{ijklm}$ & 0.5 & 1 \\
${\cal I}_{ijklmn}$ & 0.5 & 1 \\
${\cal J}_{ijklmn}$ & 0 & 1  \\
${\cal I}_{ijklmno}$ & 0 & 1 \\
${\cal I}_L (\ell > 7)$ & - & 1 \\
${\cal J}_L (\ell > 6)$ & - & 1 \\
\end{tabular}
\end{ruledtabular}
\end{table}


\section{Summary of Current Post-Newtonian Results}
\label{sec:summary-current-post}

The post-Newtonian approximation is a slow-motion, weak-field
approximation to general relativity with an expansion parameter
$\epsilon \sim (v/c)^2 \sim (Gm/rc^2)$. For a binary system of two
point masses $m_1$ and $m_2$, $v$ is the magnitude of the relative
velocity, $m = m_1 + m_2$, and $r$ is the separation.  In order to
produce a post-Newtonian waveform, it is necessary to solve both the
post-Newtonian equations of motion describing the binary, and the
post-Newtonian equations describing the generation of gravitational
waves.  For a complete review of post-Newtonian methods applied to
inspiralling compact binaries, see~\cite{Blanchet2006}.


\subsection{Post-Newtonian Computation of Radiative Multipoles}
\label{sec:post-newt-comp}

In Sec. \ref{sec:polar-wavef-modes} the spin-weighted spherical
harmonic components have been related to the radiative multipole
moments. In order to use
Eq.~(\ref{eq:ComponentsFromRadiativeMultipoles}) it is necessary to
relate the radiative moments to a description of the compact binary
system.  The post-Newtonian wave generation formalism has been
developed~\cite{Blanchet86, Blanchet88, Blanchet92, Blanchet95,
  Blanchet98a, Blanchet98b, Blanchet98} in a systematic manner to
relate the radiative multipole moments $\{{\cal U}_L,{\cal V}_L\}$ to
a set of six STF source moments $\{{\cal I}_L,{\cal J}_L,{\cal
  W}_L,{\cal X}_L,{\cal Y}_L,{\cal Z}_L\}$, which can be computed from
the stress-energy pseudo-tensor of the material and gravitational
fields of the source.  A set of two canonical source moments $\{{\cal
  M}_L,{\cal S}_L\}$ can be computed as a nonlinear functional of the
source moments as an intermediate step between the source moments and
the radiative moments.  The radiative moments are then given as
nonlinear (and even nonlocal) functionals of the canonical moments.
It turns out that two of the source moments, the source mass moments
${\cal I}_L$ and the source current moments ${\cal J}_L$ are dominant,
while the other four parameterize a gauge transformation and only make
a contribution to the canonical source moments starting at 2.5PN
order.  To the post-Newtonian order needed in this paper, only the
canonical mass quadrupole moment contains a correction from its
corresponding source moment~\cite{Blanchet96,Blanchet02}
\begin{equation}
{\cal M}_{ij} = {\cal I}_{ij} + \frac{4 G}{c^5} \left(
 {\cal W}^{(2)} {\cal I}_{ij} - {\cal W}^{(1)} {\cal I}^{(1)}_{ij} \right)
+ O(\epsilon^{7/2}),
\end{equation}
where ${\cal T}^{(p)}_L$ denotes the $p$th time derivative of ${\cal
  T}_L$ and $O(\epsilon^k)$ denotes the $k$th-order and higher
post-Newtonian corrections that are unknown.  The other canonical
source moments are simply related to the source moments
\begin{eqnarray}
{\cal M}_L &=& {\cal I}_L + O(\epsilon^{5/2}), \\
{\cal S}_L &=& {\cal J}_L + O(\epsilon^{5/2}).
\end{eqnarray}

The radiative mass quadrupole is given by
\begin{widetext}
\begin{eqnarray}
\label{eq:RadiativeMassQuadrupole}
{\cal U}_{ij}(T_R)  &=& {\cal M}_{ij}^{(2)}(T_R) + \frac{2GM}{c^3}
\int_{0}^{\infty} d\tau {\cal M}_{ij}^{(4)} (T_R - \tau) \left[
\ln{\left(\frac{c\tau}{2r_0} \right)} + \frac{11}{12} \right] -
\frac{2}{7} \frac{G}{c^5} \int_{0}^{\infty} d\tau {\cal M}_{a \langle
i}^{(3)}(T_R - \tau) {\cal M}_{j \rangle a}^{(3)}(T_R - \tau)
\nonumber \\* &&+ \frac{G}{c^5} \left\{ \frac{1}{7} {\cal M}_{a
\langle i}^{(5)} {\cal M}^{~}_{j \rangle a} - \frac{5}{7} {\cal M}_{a
\langle i}^{(4)} {\cal M}_{j \rangle a}^{(1)} - \frac{2}{7} {\cal
M}_{a \langle i}^{(3)} {\cal M}_{j \rangle a}^{(2)} + \frac{1}{3}
\epsilon^{}_{ab \langle i} {\cal M}_{j \rangle a}^{(4)} {\cal S}^{}_b
\right\} \nonumber \\* &&+ \frac{2G^2M^2}{c^6} \int_{0}^{\infty} d\tau
{\cal M}_{ij}^{(5)}(T_R - \tau) \left[ \ln^2{\left(\frac{c\tau}{2r_0}
\right)} + \frac{57}{70} \ln{\left(\frac{c\tau}{2r_0} \right)} +
\frac{124627}{44100} \right] + O(\epsilon^{7/2}).
\end{eqnarray}
\end{widetext}
The first integral in the above expression is the dominant radiation
tail at 1.5PN order obtained in~\cite{Blanchet92}.  The 2.5PN
non-linear memory integral has been obtained in~\cite{Christodoulou91,
  Wiseman91, Thorne92a, Blanchet92, Blanchet98a}.  The other
non-linear interactions at 2.5PN order were obtained
in~\cite{Blanchet98a}.  Finally the 3PN ``tail of tail'' integral was
derived in~\cite{Blanchet98b}.  The tail integrals involve nonlinear
interactions with the mass monopole $M$ (equivalent to the ADM mass)
of the system.  The tail integrals also contain a freely-specifiable
constant $r_0$ that corresponds to the choice of the origin of
radiative time $T$ with respect to harmonic time $t$, and enters the
relation between the retarded time in radiative coordinates and the
retarded time $t - r/c$ in harmonic coordinates (the coordinates in
which the equations of motion are given)~\cite{Wiseman93,Blanchet93}:
\begin{equation}
\label{eq:RadiativeTimeFromHarmonicTime}
T_R = t - \frac{r}{c} - \frac{2 G M}{c^3} \ln{\left(\frac{r}{r_0}\right)}.
\end{equation}

The remaining radiative multipole moments are given
by~\cite{Blanchet92,Blanchet98a,Blanchet98b}
\begin{widetext}
\begin{eqnarray}
{\cal U}_{ijk}(T_R) &=& {\cal M}_{ijk}^{(3)}(T_R) + 
\frac{2GM}{c^3} \int_{0}^{\infty} d\tau {\cal M}_{ijk}^{(5)} (T_R - \tau)
\left[ \ln{\left(\frac{c\tau}{2r_0} \right)} + \frac{97}{60} \right] +
O(\epsilon^{5/2}), \\
{\cal U}_{ijkl}(T_R) &=& {\cal M}_{ijkl}^{(4)}(T_R) + 
\frac{2GM}{c^3} \int_{0}^{\infty} d\tau {\cal M}_{ijkl}^{(6)} (T_R - \tau)
\left[ \ln{\left(\frac{c\tau}{2r_0} \right)} + \frac{59}{30} \right] 
+ \frac{2}{5} \frac{G}{c^3} \int_{0}^{\infty} d\tau 
{\cal M}_{\langle ij}^{(3)}(T_R - \tau) {\cal M}_{kl \rangle}^{(3)}(T_R - \tau)
\nonumber \\* &&- \frac{G}{c^3} \left\{ \frac{21}{5} 
{\cal M}_{\langle ij}^{(5)} {\cal M}^{}_{kl \rangle} +
\frac{63}{5} {\cal M}_{\langle ij}^{(4)} {\cal M}_{kl \rangle}^{(1)} +
\frac{102}{5} {\cal M}_{\langle ij}^{(3)} {\cal M}_{kl \rangle}^{(2)} 
\right\} + O(\epsilon^{5/2}), \label{eq:RadiativeMassHexadecapole}\\
{\cal U}_L &=& {\cal M}_L^{(\ell)}(T_R) + O(\epsilon^{3/2}), \\
{\cal V}_{ij}(T_R) &=& {\cal S}_{ij}^{(2)}(T_R) + 
\frac{2GM}{c^3} \int_{0}^{\infty} d\tau {\cal S}_{ij}^{(4)} (T_R - \tau)
\left[ \ln{\left(\frac{c\tau}{2r_0} \right)} + \frac{7}{6} \right] +
O(\epsilon^{5/2}), \\
{\cal V}_{ijk}(T_R) &=& {\cal S}_{ijk}^{(3)}(T_R) + 
\frac{2GM}{c^3} \int_{0}^{\infty} d\tau {\cal S}_{ijk}^{(5)} (T_R - \tau)
\left[ \ln{\left(\frac{c\tau}{2r_0} \right)} + \frac{5}{3} \right] 
\nonumber \\* && +
\frac{G}{c^3} \left\{ \frac{1}{10} \epsilon^{}_{ab \langle i} 
{\cal M}_{j\underline{a}}^{(5)} {\cal M}^{}_{k \rangle b} -
\frac{1}{2} \epsilon^{}_{ab \langle i} {\cal M}_{j\underline{a}}^{(4)} 
{\cal M}_{k \rangle b}^{(1)} - 
2 {\cal S}^{}_{\langle i} {\cal M}^{(4)}_{jk\rangle} 
\right\} + O(\epsilon^{5/2}), \\
{\cal V}_L &=& {\cal S}_L^{(\ell)}(T_R) + O(\epsilon^{3/2}).
\end{eqnarray}
\end{widetext}


\subsection{Adiabatic inspiral of quasi-circular orbits}
\label{sec:adiab-inspr-quasi}

Solving the equations of motion yields explicit expressions for the
accelerations of each body in terms of the positions and velocities of
the two bodies~\cite{Jaranowski98a, Jaranowski99a, Damour00a,
  Damour01a, Blanchet00a, Blanchet01a, Damour01b, Blanchet04, Itoh01,
  Itoh03, Itoh04}.  The two-body equations of motion can then be
reduced to relative equations of motion in the center-of-mass frame in
terms of the relative position ${\vec x}$ and velocity ${\vec
  v}$~\cite{Blanchet03a}.  The relative acceleration ${\vec a}$ is
currently known through 3.5PN order~\cite{Blanchet02a,Blanchet04a}.
The effects of radiation reaction (due to the emission of
gravitational waves) enters the relative acceleration starting at
2.5PN order.

The emission of gravitational radiation causes the orbits of an
isolated binary system to circularize~\cite{Peters1964}.  Thus for the
remainder of this paper the orbital evolution of the binary will be
modeled as a slow adiabatic inspiral of a quasi-circular orbit.  In
addition, it will be assumed that the individual compact objects are
non-spinning.

The orbital plane is chosen to be the $X$-$Y$ plane with the orbital phase
$\phi$ defining the direction of the unit vector ${\vec n} = {\vec x}/r$
along the relative separation vector by
\begin{equation}
{\bf n} = \cos{\phi}~ {\vec e_X} + \sin{\phi}~ {\vec e_Y}.
\end{equation}
Then the motion of the binary can be described using the rotating
orthonormal triad $({\vec n},{\vec \lambda},{\vec e_Z})$ with ${\vec
\lambda} = {\vec e_Z} \times {\vec n}$.

The relative position, velocity, and acceleration are given by
\begin{eqnarray}
{\vec x} &=& r {\vec n}, \\
\label{eq:EOMv}
{\vec v} &=& \dot r {\vec n} + r \omega {\vec \lambda}, \\
\label{eq:EOMa}
{\vec a} &=& (\ddot r - r \omega^2) {\vec n} + 
(r \dot \omega + 2 \dot r \omega) {\vec \lambda},
\end{eqnarray}
where the orbital frequency $\omega = \dot \phi$.  Through 2PN order,
it is possible to model the motion of the binary as a circular orbit
with the solution $\ddot r = \dot r = \dot \omega = 0$ and $r \omega^2
= - {\vec n} \cdot {\vec a}$.

At 2.5PN order, however, the inspiral motion must be taken into
account.  The leading order contribution to the inspiral of a
quasi-circular orbit can be obtained by examining the Newtonian
orbital energy of a circular orbit,
\begin{equation}
E = - \frac{1}{2} \nu \frac{Gm^2}{r} + O(\epsilon),
\end{equation}
where $\nu = m_1 m_2 / m^2$, and the leading order gravitational
luminosity from a circular orbit,
\begin{equation}
{\cal L} = \frac{32}{5} \nu^2 \frac{G^4 m^5}{r^5 c^5} + O(\epsilon),
\end{equation}
and assuming that the energy radiated by the gravitational waves is balanced by
the change in the orbital energy (i.e. $dE/dt = - {\cal L}$).  Then
\begin{equation}
\label{eq:rDot}
\dot r = (\frac{dE}{dt} / \frac{dE}{dr}) 
= - \frac{64}{5} \nu \frac{G^3 m^3}{r^3 c^5} + O(\epsilon^{7/2}),
\end{equation}
and similarly the orbital frequency changes by
\begin{equation}
\label{eq:OmegaDot}
\dot \omega = (\frac{d \omega}{dr} / \frac{dr}{dt})
= \frac{96}{5} \nu 
\left( \frac{G^7 m^7}{c^{10} r^{11}} \right)^{1/2} + O(\epsilon^{7/2}).
\end{equation}

Substituting (\ref{eq:rDot}) and (\ref{eq:OmegaDot}) into
Eqs.~(\ref{eq:EOMv}) and~(\ref{eq:EOMa}) and noting that $\ddot r =
O(\epsilon^5)$, the following expressions for the 3PN inspiral
relative velocity and relative acceleration are obtained:
\begin{eqnarray}
\label{eq:V3PN}
{\vec v} &=& r \omega {\vec \lambda} 
- \frac{64}{5} \nu \frac{G^3 m^3}{r^3 c^5} {\vec n} + O(\epsilon^{7/2}), \\
\label{eq:a3PN}
{\vec a} &=& -\omega^2 {\vec x} 
- \frac{32}{5} \nu \frac{G^3 m^3}{c^5 r^4} {\vec v} + O(\epsilon^{7/2}).
\end{eqnarray}
where the 3PN orbital angular frequency is~\cite{Blanchet00a,Blanchet01a}
\begin{eqnarray}
\label{eq:w3PN}
\omega^2 &=& \frac{Gm}{r^3} \left\{ 1 + \gamma \left(-3 + \nu \right) + 
\gamma^2 \left( 6 + \frac{41}{4} \nu + \nu^2 \right) 
\right. \nonumber \\*  &&+\left.  
\gamma^3 \left( -10 + \left[ -\frac{75707}{840} + \frac{41}{64} \pi^2 + 
22 \ln{\left( \frac{r}{r'_{0}} \right)} \right] \nu \right. \right.
\nonumber \\* &&+ \left. \left. \frac{19}{2} \nu^2 
+ \nu^3 \right) \right\}  + O(\epsilon^4),
\end{eqnarray}
where the post-Newtonian parameter
\begin{equation}
\gamma \equiv \frac{Gm}{rc^2},
\end{equation}
and $r'_{0}$ is a gauge constant.

As $\gamma$ is defined with respect to the harmonic-coordinate separation $r$,
it is convenient to introduce the frequency-related post-Newtonian parameter
\begin{equation}
x \equiv \left( \frac{Gm\omega}{c^3} \right)^{2/3}.
\end{equation}
Inverting (\ref{eq:w3PN}) order by order, yields
\begin{eqnarray}
\gamma &=& x \left\{ 1 + x \left( 1 - \frac{1}{3} \nu \right) + 
x^2 \left( 1 - \frac{65}{12} \nu \right) \right. \nonumber \\* &&+ \left.
x^3 \left( 1 + \left[ - \frac{2203}{2520} - \frac{41}{192} \pi^2 - \frac{22}{3}
\ln{\left( \frac{r}{r_{0}^{'}} \right)} \right] \nu 
\right. \right. \nonumber \\* &&+ \left. \left. \frac{229}{36} \nu^2
+ \frac{1}{81} \nu^3 \right) \right\} + O(\epsilon^4).
\end{eqnarray}

\subsection{Source Multipole Moments}
\label{sec:source-mult-moments}

Defining $\delta m = m_1 - m_2$, the source mass multipole moments for
circular orbits are given
by~\cite{Blanchet89,Blanchet95a,Blanchet02}
\begin{eqnarray}
{\cal I} &=& m \left\{ 1 - \frac{1}{2} \nu \gamma + 
 \gamma^2 \frac{1}{8} \nu \left( 7 - \nu \right) \right\} + O(\epsilon^3), \\
{\cal I}_i &=& 0,
\end{eqnarray}
\begin{widetext}
\begin{eqnarray}
\label{eq:SourceMassQuadrupole}
{\cal I}_{ij} &=& \nu m \left\{ \left[ 1 - 
\gamma \left( \frac{1}{42} + \frac{13}{14} \nu \right) - 
\gamma^2 \left( \frac{461}{1512} + \frac{18395}{1512} \nu + 
\frac{241}{1512} \nu^2 \right)  
\right. \right. \nonumber \\* &&+ \left. \left. 
\gamma^3 \left( \frac{395899}{13200} - 
\frac{428}{105} \ln{\left( \frac{r}{r_0} \right)} +
\left[ \frac{3304319}{166320} -
\frac{44}{3} \ln{\left( \frac{r}{r'_{0}} \right)}\right] \nu +
\frac{162539}{16632} \nu^2 + \frac{2351}{33264} \nu^3 \right) \right] x_{<ij>} 
\right. \nonumber \\* &&+ \left.
\left[ \frac{11}{21} ( 1 - 3 \nu) +
\gamma \left( \frac{1607}{378} - \frac{1681}{378} \nu + 
\frac{229}{378} \nu^2 \right) +
\gamma^2 \left( 
\frac{428}{105} \ln{\left( \frac{r}{r_0} \right)} - \frac{357761}{19800} -
\frac{92339}{5544} \nu \right. 
\right. \right. \nonumber \\* &&+ \left. \left. \left. 
\frac{35759}{924} \nu^2 + 
\frac{457}{5544} \nu^3 \right) \right] \frac{r^2}{c^2} v_{<ij>} 
+ \frac{48}{7}\frac{G^2 m^2 \nu}{r c^5} x_{<i} v_{j>} \right\} 
+ O(\epsilon^{7/2}), \\
{\cal I}_{ijk} &=& - \nu \delta m \left\{ \left[ 1 - \gamma \nu -
\gamma^2 \left( \frac{139}{330} + \frac{11923}{660} \nu + 
\frac{29}{110} \nu^2 \right) \right] x_{<ijk>} \right.
\nonumber \\* && \left. +
\left[ 1 - 2 \nu + 
\gamma \left( \frac{1066}{165} - \frac{1433}{330} \nu + 
\frac{21}{55} \nu^2 \right) \right] \frac{r^2}{c^2} x_{<i} v_{jk>} 
\right\} + O(\epsilon^{5/2}), \\
{\cal I}_{ijkl} &=& \nu m \left\{ \left[ 1 - 3 \nu +
\gamma \left( \frac{3}{110} - \frac{25}{22} \nu + \frac{69}{22} \nu^2 \right) -
\gamma^2 \left( \frac{126901}{200200} + \frac{58101}{2600} \nu - 
\frac{204153}{2860} \nu^2 - \frac{1149}{1144} \nu^3 \right) \right] x_{<ijkl>} 
\right. \nonumber \\* && \left. + ~
\left[ \frac{78}{55} \left( 1 - 5 \nu + 5 \nu^2 \right) + 
\gamma \left( \frac{30583}{3575} - \frac{107039}{3575} \nu + 
\frac{8792}{715} \nu^2 - \frac{639}{715} \nu^3 \right) \right] 
\frac{r^2}{c^2} x_{<ij} v_{kl>} 
\right. \nonumber \\* && \left. + ~
\left[ \frac{71}{715} \left( 1 - 7 \nu + 14 \nu^2 - 7 \nu^3 \right) \right] 
\frac{r^4}{c^4} v_{<ijkl>} \right\} + O(\epsilon^{5/2}), \\
{\cal I}_{ijklm} &=& - \nu \delta m \left\{ \left[
1 - 2 \nu + 
\gamma \left( \frac{2}{39} - \frac{47}{39} \nu + \frac{28}{13} \nu^2\right) 
\right] x_{<ijklm>} +
\frac{70}{39} \left( 1 - 4 \nu + 3 \nu^2 \right) 
\frac{r^2}{c^2} x_{<ijk} v_{lm>} \right\} + O(\epsilon^2), \\
{\cal I}_{ijklmn} &=& \nu m \left\{ \left[
1 - 5 \nu + 5 \nu^2 +
\gamma \left( \frac{1}{14} - \frac{3}{2} \nu + 6 \nu^2 - \frac{11}{2} \nu^3
\right) \right] x_{<ijklmn>} \right .
\nonumber \\* && \left. +
\frac{15}{7} \left( 1 - 7 \nu + 14 \nu^2 - 7 \nu^3 \right) 
\frac{r^2}{c^2} x_{<ijkl} v_{mn>} \right\} + O(\epsilon^2),\\
{\cal I}_{ijklmno} &=& - \nu \delta m \left\{ \left[
1 - 4 \nu + 3 \nu^2 + 
\gamma \left( \frac{3}{34} - \frac{26}{17} \nu + \frac{83}{17} \nu^2
- \frac{57}{17} \nu^3 \right) \right] x_{<ijklmno>} \right .
\nonumber \\* && \left. +
\frac{42}{17} \left( 1 - 6 \nu + 10 \nu^2 - 4 \nu^3 \right) 
\frac{r^2}{c^2} x_{<ijklm} v_{no>} \right\} + O(\epsilon^2),
\end{eqnarray}
\begin{eqnarray}
{\cal I}_{ijklmnop} &=& \nu m \left\{  \left[
1 - 7 \nu + 14 \nu^2 - 7 \nu^3 +
\gamma \left( \frac{35}{342} - \frac{73}{38} \nu + \frac{317}{38} \nu^2 
- \frac{973}{57} \nu^3 + \frac{301}{38} \nu^4 \right) \right] x_{<ijklmnop>} 
\right. \nonumber \\* && \left. +
\frac{476}{171} \left( 1 - 9 \nu + 27 \nu^2 - 30 \nu^3 + 9 \nu^4\right) 
\frac{r^2}{c^2} x_{<ijklmn} v_{op>} \right\} + O(\epsilon^2),
\end{eqnarray}
through $\ell = 8$.  In Appendix~\ref{sec:1pn-source-mass} an expression is given 
for the source mass moments to 1PN order for arbitrary $\ell$.

The source current multipole moments for circular orbits
are~\cite{Damour:1990ji,Blanchet95,Blanchet02}
\begin{eqnarray}
{\cal J}_i &=& L_i + O(\epsilon), \\
{\cal J}_{ij} &=& - \frac{\delta m}{m} \left[ 1 + 
\gamma \left( \frac{67}{28} - \frac{2}{7} \nu \right) +
\gamma^2 \left( \frac{13}{9} - \frac{4651}{252} \nu - 
\frac{1}{168} \nu^2 \right) \right] L_{<i} x_{j>} + O(\epsilon^{5/2}), \\
{\cal J}_{ijk} &=& \left[ 1 - 3 \nu +
\gamma \left( \frac{181}{90} - \frac{109}{18} \nu + \frac{13}{18} \nu^2 \right)
\right] L_{<i} x_{jk>} + \frac{7}{45} \left( 1 - 5 \nu + 5 \nu^2
\right) \frac{r^2}{c^2} L_{<i} v_{jk>} + O(\epsilon^2), \\ 
{\cal J}_{ijkl} &=& - \frac{\delta m}{m} \left\{ \left[ 1 - 2 \nu +
\gamma \left( \frac{20}{11} - \frac{155}{44} \nu + \frac{5}{11} \nu^2 \right)
\right] L_{<i} x_{jkl>} + \frac{4}{11} \left( 1 - 4 \nu + 3 \nu^2
\right) \frac{r^2}{c^2} L_{<i} x_j v_{kl>} \right\} + O(\epsilon^2), \\ 
{\cal J}_{ijklm} &=& \left[ 1 - 5 \nu + 5 \nu^2 +
\gamma \left( \frac{1549}{910} - \frac{1081}{130} \nu + \frac{107}{13} \nu^2 - 
\frac{29}{26} \nu^3 \right)
\right] L_{<i} x_{jklm>} \nonumber \\* && + 
\frac{54}{91} \left( 1 - 7 \nu + 14 \nu^2 - 7 \nu^3
\right) \frac{r^2}{c^2} L_{<i} x_{jk} v_{lm>} + O(\epsilon^2), \\ 
{\cal J}_{ijklmn} &=& - \frac{\delta m}{m} \left\{ 1 - 4 \nu + 3 \nu^2
\right\} L_{<i} x_{jklmn>}  + O(\epsilon), \\ 
{\cal J}_{ijklmno} &=& \left\{ 1 - 7 \nu + 14 \nu^2 - 7 \nu^3
\right\} L_{<i} x_{jklmno>}  + O(\epsilon), \\ 
{\cal J}_{ijklmnop} &=& - \frac{\delta m}{m} \left\{ 1 - 6 \nu + 10 \nu^2
- 4 \nu^3 \right\} L_{<i} x_{jklmnop>}  + O(\epsilon),
\end{eqnarray}
\end{widetext}
where $L_i = \nu m \epsilon_{iab} x_a v_b$.  In
Appendix~\ref{sec:0pn-source-current} an expression is given for the
source current moments to leading order for arbitrary $\ell$.

Finally, the gauge monopolar moment ${\cal W}$ for a circular orbit is 
\begin{equation}
{\cal W} = \frac{1}{3} \mu x_a v_a,
\end{equation}
which is proportional to ${\dot r} \sim O(\epsilon^{5/2})$.  As ${\cal
W}$ is already a 2.5PN correction to the source mass quadrupole, it
gives no contribution in the present case, and the canonical moments
will simply be given by the source moments.


\section{Results}
\label{sec:results}

The computation of the spin-weighted spherical harmonic components is
now an exercise in algebra.  The evaluation of the tail, memory, and
``tail of tail'' integrals requires special treatment that is
described in Sec.~\ref{sec:deal-with-hered}.  The spin-weighted
spherical harmonic components are listed in
Sec.~\ref{sec:spin-weight-spher}.  They will contain terms that depend
upon the freely-specifiable constant $r_0$ (see the discussion between
Eqs.~(\ref{eq:RadiativeMassQuadrupole}) and
(\ref{eq:RadiativeTimeFromHarmonicTime})).  These terms can be
absorbed into a redefinition of the phase as will be discussed in
Sec.~\ref{sec:absorb-ampl-terms}.


\subsection{Evaluation of hereditary terms}
\label{sec:deal-with-hered}

The hereditary terms in the radiative multipole moments involve
integrals over the entire past history of the binary.  These integrals
fall into two types, the tail integrals (including the ``tail of
tail'' term) that have logarithmic terms and physically correspond to
the backscattering of the gravitational waves off the background
curvature~\cite{Blanchet92}; and the memory integrals that can
physically be thought of as the re-radiation of the stress-energy of
the propagating waves~\cite{Christodoulou91,Wiseman91,Thorne92a}.

Inserting the canonical multipole moments into the tail integrals of
the radiative multipole moments yields expressions of the form
\begin{equation}
I_1(T_R) = \int_0^\infty d\tau F(T_R-\tau) e^{-ik\omega(T_R-\tau)} 
\left[ \ln{\left(\frac{c\tau}{2r_0} \right)} + b \right] ,
\end{equation}
where $k$ is an integer (the index $m$ of the $h^{\ell m}$ being
computed), $b$ is a rational number, and $F(T_R-\tau)$ represents a
function whose time dependence enters only through its dependence on
the orbital frequency.  In~\cite{Blanchet93,Arun2004}, it has been
shown that the oscillatory term in the integrand combined with the
slow adiabatic evolution of the orbital frequency allow the integral
to be approximated by using the orbital frequency of a fixed circular
orbit at the current value of $T_R$, so that
\begin{eqnarray}
\label{eq:TailApprox}
I_1(T_R) &=& F(T_R) e^{-ik\phi} \left\{ \int_0^\infty d\tau e^{ik\omega \tau}
\left[ \ln{\left(\frac{c\tau}{2r_0} \right)} + b \right]
\right. \nonumber \\* &&+  \left. O(\xi \ln \xi) \right\},
\end{eqnarray}
where $\xi \sim {\dot \omega}/\omega^2$, is the adiabatic parameter
describing the decay of the orbit which is $O(\epsilon^{5/2})$.
Equation~(\ref{eq:TailApprox}) can be evaluated using the identity
\begin{equation}
\label{eq:LogIntegral}
\int_0^\infty d\tau \ln{\tau} e^{-\sigma \tau} = 
- \frac{1}{\sigma} \left( \gamma_E + \ln{\sigma} \right), 
\end{equation}
where $\gamma_E$ is Euler's constant.  Using (\ref{eq:LogIntegral}) to
evaluate (\ref{eq:TailApprox}) yields
\begin{eqnarray}
I_1(T_R) &=& -\frac{1}{k \omega} F(T_R) e^{-ik\phi} 
\left\{ \frac{\pi}{2} \right.
\nonumber \\* &&+ \left. i \left[ \gamma_E + \ln{\left( \frac{2 k \omega
r_0}{c} \right)} - b \right] \right\} .
\end{eqnarray}
For reasons that will be explained in Sec.~\ref{sec:absorb-ampl-terms}, it
is convenient to express the above result as
\begin{eqnarray}
I_1(T_R) &=& -\frac{1}{k \omega} F(T_R)e^{-ik\phi} \left\{
\frac{\pi}{2} \right.
\nonumber \\* &&+ \left. i \left[ \frac{3}{2} \ln{\left(\frac{x}{x_0}\right)} +
\ln{\frac{k}{2}} + \frac{11}{12} - b \right] \right\},
\end{eqnarray}
where
\begin{equation}
\label{eq:LogParameter}
\ln{x_0} \equiv \frac{11}{18} - \frac{2}{3} \gamma_E - \frac{4}{3} \ln{2}
+ \frac{2}{3} \ln{\left(\frac{G m}{c^2 r_0} \right)}.
\end{equation}

A similar argument holds for the ``tail of tail'' integral in
Eq.~(\ref{eq:RadiativeMassQuadrupole}) where a term of the form
\begin{equation}
I_2(T_R) = \int_0^\infty d\tau  F(T_R-\tau) e^{-ik\omega(T_R-\tau)} 
\left[ \ln{\left(\frac{c\tau}{2r_0} \right)}\right]^2
\end{equation}
is found.  In this case the integral can be evaluated with the aid of
the identity
\begin{equation}
\int_0^\infty d\tau (\ln{\tau})^2 e^{-\sigma \tau} = 
\frac{1}{\sigma} \left[ \frac{\pi^2}{6} + (\gamma_E + \ln{\sigma})^2 \right],
\end{equation}
to yield
\begin{eqnarray}
{I}_{2}(T_R) &=& \frac{1}{k \omega} F(T_R)e^{-ik\phi} \left\{
\pi \left[ \frac{3}{2} \ln{\left(\frac{x}{x_0}\right)} +
\ln{\frac{k}{2}} + \frac{11}{12} \right] \right.
\nonumber \\* &&+ \left. i \left[ \frac{3}{2} \ln{\left(\frac{x}{x_0}\right)} +
\ln{\frac{k}{2}} + \frac{11}{12} \right]^2 \right\}.
\end{eqnarray}

The memory integral in Eq.~(\ref{eq:RadiativeMassHexadecapole}) leads to
an integral of the form
\begin{equation}
I_3(T_R) = \int_0^\infty  F(T_R-\tau) e^{-ik\omega(T_R-\tau)} d\tau.
\end{equation}
Using similar arguments as for the tail integrals, Ref.~\cite{Arun2004} has
shown that this integral can be approximated by
\begin{equation}
I_3(T_R) = \frac{i}{k \omega} F(T_R) e^{-ik\phi} \left[ 1 + O(\xi) \right] .
\end{equation}

The memory integrals also lead to an integral of the form
\begin{equation}
I_4(T_R) = \int_0^\infty x^5(T_R-\tau) d\tau.
\end{equation}
Unlike the other integrals, it does not depend upon the orbital phase,
and thus is much more sensitive to the past history of the binary.
This is the non-linear memory effect described
in~\cite{Christodoulou91, Wiseman91, Thorne92a}.  This integral has
been evaluated by~\cite{Wiseman91,Arun2004} using the evolution of the
frequency parameter $x$ found by integrating Eq.~(\ref{eq:OmegaDot})
\begin{equation}
x(t) = \frac{1}{4} \left[ \left( \frac{\nu c^3}{5 G m} \right)
\left( t_c - t \right) \right]^{-1/4} \left[ 1 + O(\epsilon) \right]
\end{equation}
where $t_c$ denotes the time of coalescence.  With this model, 
\begin{equation}
I_4(T_R) = \frac{5}{64} \frac{G m}{\nu c^3} x \left[ 1 + O(\epsilon) \right].
\end{equation}
As discussed in~\cite{Blanchet92,Wiseman91,Arun2004}, this contribution
to the waveform will be very difficult to detect as it is essentially
DC in character corresponding to a steadily-growing part of the
waveform with no dependence on the orbital phase.  It does, however,
build up during the entire inspiral which leads to its magnitude being
comparable with the 0PN quadrupolar term despite arising formally as a
2.5PN contribution. See~\cite{Thorne92a} for a discussion on
strategies for detecting the memory piece of the waveform.


\subsection{Spin-weighted spherical harmonic components}
\label{sec:spin-weight-spher}

Examining Eq.~(\ref{eq:ComponentsFromRadiativeMultipoles}), seems to
show that the individual spin-weighted spherical harmonic components
could obtain contributions from both mass and current radiative
multipoles.  It turns out for the non-spinning case, however, that the
mass (current) multipoles only contribute to components with $\ell +
m$ even (odd).  Because of this separation, and since~\cite{thorne80}
$U^{\ell m \ast} = (-1)^m U^{\ell,-m}$ and $V^{\ell m \ast} = (-1)^m
V^{\ell,-m}$, it follows from
Eq.~(\ref{eq:ComponentsFromRadiativeMultipoles}) that
\begin{equation}
h^{\ell,-m} =
\begin{cases}
(-1)^m h^{\ell m \ast} & \text{($\ell + m$ even)} \\
(-1)^{m+1} h^{\ell m \ast} & \text{($\ell + m$ odd)} \\
 \end{cases}
\end{equation}
which simply reduces to
\begin{equation}
\label{eq:mNegative}
h^{\ell,-m} = (-1)^\ell h^{\ell m \ast}
\end{equation}

Substituting the source mass quadrupole
(\ref{eq:SourceMassQuadrupole}) into the expression for the radiative
mass quadrupole (\ref{eq:RadiativeMassQuadrupole}), taking the
appropriate time derivatives, substituting the equations of motion
(\ref{eq:a3PN}), and evaluating the hereditary integrals using the
techniques described in Sec.~\ref{sec:deal-with-hered}, the dominant
component of the waveform is given to 3PN order as
\begin{widetext}
\begin{eqnarray}
\label{eq:22-mode}
h^{22} =&& -8 \sqrt{\frac{\pi }{5}} \frac{G \nu m}{c^2 R} e^{-2 i \phi } x
\left\{ 1 - x \left( \frac{107}{42} - \frac{55}{42} \nu \right)
+  x^{3/2} \left[ 2 \pi + 6 i \ln{\left(\frac{x}{x_0}\right)} \right]
-  x^2 \left( \frac{2173}{1512} + \frac{1069}{216} \nu - 
\frac{2047}{1512} \nu^2 \right)
\right. \nonumber \\* &&- \left. x^{5/2}
\left[ \left(\frac{107}{21} - \frac{34}{21} \nu \right) \pi + 24 i \nu +
\left(\frac{107 i}{7} - \frac{34 i}{7} \nu \right) 
\ln{\left(\frac{x}{x_0}\right)} \right]  +
x^3 \left[ \frac{27027409}{646800} - \frac{856}{105} \gamma_E 
+ \frac{2}{3} \pi^2 - \frac{1712}{105} \ln{2} 
\right. \right. \nonumber \\* && \left. \left. - \frac{428}{105} \ln{x} 
- 18 \left[ \ln{\left(\frac{x}{x_0}\right)} \right]^2
- \left( \frac{278185}{33264} - \frac{41}{96} \pi^2 \right) \nu
- \frac{20261}{2772} \nu^2 + \frac{114635}{99792} \nu^3 + \frac{428 i}{105} \pi
+ 12 i \pi \ln{\left(\frac{x}{x_0}\right)} \right] \right.\nonumber \\* &&
\left. + ~ O(\epsilon^{7/2}) \right\}, 
\end{eqnarray}
where the constant $r_0$ has been eliminating from the expression by
using Eq.~(\ref{eq:LogParameter}).  The remaining components are given
by
\begin{eqnarray}
h^{21} &=& -\frac{8 i}{3} \sqrt{\frac{\pi }{5}} \frac{G \nu \delta m}{c^2 R} 
e^{-i \phi } x^{3/2} \left\{ 1 - 
x \left( \frac{17}{28} - \frac{5}{7} \nu \right) +
x^{3/2} \left[ \pi - \frac{i}{2} - 2 i \ln{2} + 
3 i \ln{\left(\frac{x}{x_0}\right)}
\right] \right. \nonumber \\* &&- \left. 
x^2 \left( \frac{43}{126} + \frac{509}{126} \nu - \frac{79}{168} \nu^2 \right) 
 + O(\epsilon^{5/2}) \right\}, \\
h^{20} &=& \frac{2}{7} \sqrt{\frac{10 \pi}{3}} \frac{G \nu m}{c^2 R} x
\left[ 1 + O(\epsilon) \right], \\
h^{33} &=& 3 i \sqrt{\frac{6 \pi }{7}} \frac{G \nu \delta m}{c^2 R} 
e^{-3 i \phi } x^{3/2} \left\{ 1 - x \left( 4 - 2 \nu \right) +
x^{3/2} \left[  3 \pi - \frac{21 i}{5} + 6 i \ln{\frac{3}{2}} + 
9 i \ln{\left(\frac{x}{x_0}\right)} \right] 
\right. \nonumber \\* &&+ \left. x^2 \left( \frac{123}{110} - 
\frac{1838}{165} \nu + \frac{887}{330} \nu^2 \right)  + O(\epsilon^{5/2}) 
\right\}, \\
h^{32} &=& -\frac{8}{3} \sqrt{\frac{\pi }{7}} \frac{G \nu m}{c^2 R} 
e^{-2 i \phi } x^2 \left\{ 1 - 3 \nu - x \left( \frac{193}{90} - 
\frac{145}{18} \nu + \frac{73}{18} \nu^2 \right) \right. \nonumber \\*
&&+ \left.
x^{3/2} \left[  2 \pi (1  - 3 \nu) - 3 i + \frac{66 i}{5} \nu +  
6 i (1 - 3 \nu) \ln{\left(\frac{x}{x_0}\right)} \right] 
 + O(\epsilon^2) \right\}, \\
h^{31} &=& - \frac{i}{3} \sqrt{\frac{2 \pi }{35}} \frac{G \nu \delta m}{c^2 R} 
e^{-i \phi } x^{3/2} \left\{ 1 - x \left( \frac{8}{3} + 
\frac{2}{3} \nu \right) +
x^{3/2} \left[  \pi - \frac{7 i}{5} - 2 i \ln{2} + 
3 i \ln{\left(\frac{x}{x_0}\right)} \right] 
\right. \nonumber \\* &&+ \left. x^2 \left( \frac{607}{198} - 
\frac{136}{99} \nu - \frac{247}{198} \nu^2 \right)   + O(\epsilon^{5/2}) 
\right\}, \\
h^{30} &=& \frac{16 i}{5} \sqrt{\frac{6 \pi }{35}} \frac{G \nu m}{c^2 R} 
\nu x^{7/2} [ 1 + O(\epsilon)], \\
h^{44} &=& \frac{64}{9} \sqrt{\frac{\pi }{7}} \frac{G \nu m}{c^2 R} 
e^{-4 i \phi } x^2 \left\{ 1 - 3 \nu - x \left( \frac{593}{110} - 
\frac{1273}{66} \nu + \frac{175}{22} \nu^2 \right) \right. \nonumber \\*
&&+ \left.
x^{3/2} \left[  4 \pi (1  - 3 \nu) - \frac{42 i}{5} + \frac{1193 i}{40} \nu +  
8 i (1 - 3 \nu) \ln{2} +
12 i (1 - 3 \nu) \ln{\left(\frac{x}{x_0}\right)} \right] 
\right. \nonumber \\* &&+ \left.
x^2 \left( \frac{1068671}{200200} - \frac{1088119}{28600} \nu + 
\frac{146879}{2340} \nu^2 - \frac{226097}{17160} \nu^3 \right) 
 + O(\epsilon^{5/2}) \right\}, \\
h^{43} &=& \frac{9 i}{5} \sqrt{\frac{2 \pi }{7}} \frac{G \nu \delta m}{c^2 R} 
e^{-3 i \phi } x^{5/2} \left\{ 1 - 2 \nu - x \left( \frac{39}{11} - 
\frac{1267}{132} \nu + \frac{131}{33} \nu^2 \right)  + O(\epsilon^{3/2}) 
\right\}, \\
h^{42} &=& -\frac{8}{63} \sqrt{\pi} \frac{G \nu m}{c^2 R} 
e^{-2 i \phi } x^2 \left\{ 1 - 3 \nu - x \left( \frac{437}{110} - 
\frac{805}{66} \nu + \frac{19}{22} \nu^2 \right) \right. \nonumber \\*
&&+ \left.
x^{3/2} \left[  2 \pi (1  - 3 \nu) - \frac{21 i}{5} + \frac{84 i}{5} \nu +
6 i (1 - 3 \nu) \ln{\left(\frac{x}{x_0}\right)} \right] 
\right. \nonumber \\* &&+ \left.
x^2 \left( \frac{1038039}{200200} - \frac{606751}{28600} \nu + 
\frac{400453}{25740} \nu^2 + \frac{25783}{17160} \nu^3 \right) 
 + O(\epsilon^{5/2}) \right\}, \\
h^{41} &=& - \frac{i}{105} \sqrt{2 \pi} \frac{G \nu \delta m}{c^2 R} 
e^{-i \phi } x^{5/2} \left\{ 1 - 2 \nu - x \left( \frac{101}{33} - 
\frac{337}{44} \nu + \frac{83}{33} \nu^2 \right)  + O(\epsilon^{3/2}) 
\right\},
\end{eqnarray}
\begin{eqnarray}
h^{40} &=& \frac{1}{63} \sqrt{\frac{\pi}{10}} \frac{G \nu m}{c^2 R} x
\left[ 1 + O(\epsilon) \right], \\
h^{55} &=& - \frac{125 i}{12} \sqrt{\frac{5 \pi}{66}} 
\frac{G \nu \delta m}{c^2 R} 
e^{-5 i \phi } x^{5/2} \left\{ 1 - 2 \nu - x \left( \frac{263}{39} - 
\frac{688}{39} \nu + \frac{256}{39} \nu^2 \right)  + O(\epsilon^{3/2}) 
\right\},\\
h^{54} &=& \frac{256}{45} \sqrt{\frac{\pi }{33}} \frac{G \nu m}{c^2 R} 
e^{-4 i \phi } x^3 \left\{ 1 - 5 \nu + 5 \nu^2 - x \left( \frac{4451}{910} - 
\frac{3619}{130} \nu + \frac{521}{13} \nu^2 - \frac{339}{26} \nu^3 \right)  
+ O(\epsilon^{3/2}) \right\}, \\
h^{53} &=& \frac{9 i}{20} \sqrt{\frac{3 \pi}{22}} 
\frac{G \nu \delta m}{c^2 R} 
e^{-3 i \phi } x^{5/2} \left\{ 1 - 2 \nu - x \left( \frac{69}{13} - 
\frac{464}{39} \nu + \frac{88}{39} \nu^2 \right)  + O(\epsilon^{3/2}) 
\right\},\\
h^{52} &=& - \frac{16}{135} \sqrt{\frac{\pi }{11}} \frac{G \nu m}{c^2 R} 
e^{-2 i \phi } x^3 \left\{ 1 - 5 \nu + 5 \nu^2 - x \left( \frac{3911}{910} - 
\frac{3079}{130} \nu + \frac{413}{13} \nu^2 - \frac{231}{26} \nu^3 \right)  
+ O(\epsilon^{3/2}) 
\right\}, \\
h^{51} &=& - \frac{i}{180} \sqrt{\frac{\pi}{77}} 
\frac{G \nu \delta m}{c^2 R} 
e^{-i \phi } x^{5/2} \left\{ 1 - 2 \nu - x \left( \frac{179}{39} - 
\frac{352}{39} \nu + \frac{4}{39} \nu^2 \right)  + O(\epsilon^{3/2}) 
\right\}, \\
h^{66} &=& - \frac{432}{5} \sqrt{\frac{\pi }{715}} \frac{G \nu m}{c^2 R} 
e^{-6 i \phi } x^3 \left\{ 1 - 5 \nu + 5 \nu^2 - x \left( \frac{113}{14} - 
\frac{91}{2} \nu + 64 \nu^2 - \frac{39}{2} \nu^3 \right)  + O(\epsilon^{3/2}) 
\right\}, \\
h^{65} &=& - \frac{625 i}{63} \sqrt{\frac{5 \pi}{429}} 
\frac{G \nu \delta m}{c^2 R} 
e^{-5 i \phi } x^{7/2} \left\{ 1 - 4 \nu + 3 \nu^2 
  + O(\epsilon) \right\}, \\
h^{64} &=& \frac{1024}{495} \sqrt{\frac{2 \pi }{195}} \frac{G \nu m}{c^2 R} 
e^{-4 i \phi } x^3 \left\{ 1 - 5 \nu + 5 \nu^2 - x \left( \frac{93}{14} - 
\frac{71}{2} \nu + 44 \nu^2 - \frac{19}{2} \nu^3 \right)  + O(\epsilon^{3/2}) 
\right\}, \\
h^{63} &=& \frac{81 i}{385} \sqrt{\frac{\pi}{13}} 
\frac{G \nu \delta m}{c^2 R} 
e^{-3 i \phi } x^{7/2} \left\{ 1 - 4 \nu + 3 \nu^2 
  + O(\epsilon) \right\}, \\
h^{62} &=& - \frac{16}{1485} \sqrt{\frac{\pi }{13}} \frac{G \nu m}{c^2 R} 
e^{-2 i \phi } x^3 \left\{ 1 - 5 \nu + 5 \nu^2 - x \left( \frac{81}{14} - 
\frac{59}{2} \nu + 32 \nu^2 - \frac{7}{2} \nu^3 \right)  + O(\epsilon^{3/2}) 
\right\}, \\
h^{61} &=& - \frac{i}{2079} \sqrt{\frac{2 \pi}{65}} 
\frac{G \nu \delta m}{c^2 R} 
e^{-i \phi } x^{7/2} \left\{ 1 - 4 \nu + 3 \nu^2 
  + O(\epsilon) \right\}, \\
h^{77} &=& \frac{16807 i}{180} \sqrt{\frac{7 \pi}{4290}} 
\frac{G \nu \delta m}{c^2 R} 
e^{-7 i \phi } x^{7/2} \left\{ 1 - 4 \nu + 3 \nu^2 - x \left( \frac{319}{34} - 
\frac{2225}{51} \nu + \frac{2558}{51} \nu^2 - \frac{230}{17} \nu^3 \right)
  + O(\epsilon^{3/2}) \right\},\\
h^{76} &=& - \frac{648}{35} \sqrt{\frac{3 \pi }{715}} 
\frac{G \nu m}{c^2 R} e^{-6 i \phi } x^4 \left\{ 1 - 7 \nu + 14 \nu^2 - 7 \nu^3
+ O(\epsilon) 
\right\}, \\
h^{75} &=& - \frac{3125 i}{3276} \sqrt{\frac{5 \pi}{66}} 
\frac{G \nu \delta m}{c^2 R} 
e^{-5 i \phi } x^{7/2} \left\{ 1 - 4 \nu + 3 \nu^2 - x \left( \frac{271}{34} - 
\frac{1793}{51} \nu + \frac{1838}{51} \nu^2 - \frac{134}{17} \nu^3 \right)
  + O(\epsilon^{3/2}) \right\},\\
h^{74} &=& \frac{1024}{1365} \sqrt{\frac{2 \pi }{165}} 
\frac{G \nu m}{c^2 R} e^{-4 i \phi } x^4 \left\{ 1 - 7 \nu + 14 \nu^2 - 7 \nu^3
+ O(\epsilon) 
\right\}, \\
h^{73} &=& \frac{243 i}{20020} \sqrt{\frac{3 \pi}{10}} 
\frac{G \nu \delta m}{c^2 R} 
e^{-3 i \phi } x^{7/2} \left\{ 1 - 4 \nu + 3 \nu^2 - x \left( \frac{239}{34} - 
\frac{1505}{51} \nu + \frac{1358}{51} \nu^2 - \frac{70}{17} \nu^3 
\right)  + O(\epsilon^{3/2}) \right\},\\
h^{72} &=& - \frac{8}{3003} \sqrt{\frac{\pi }{15}} 
\frac{G \nu m}{c^2 R} e^{-2 i \phi } x^4 \left\{ 1 - 7 \nu + 14 \nu^2 - 7 \nu^3
+ O(\epsilon) 
\right\}, \\
h^{71} &=& - \frac{i}{108108} \sqrt{\frac{\pi}{10}} 
\frac{G \nu \delta m}{c^2 R} 
e^{-i \phi } x^{7/2} \left\{ 1 - 4 \nu + 3 \nu^2 - x \left( \frac{223}{34} - 
\frac{1361}{51} \nu + \frac{1118}{51} \nu^2 - \frac{38}{17} \nu^3 \right)
  + O(\epsilon^{3/2}) \right\},\\
h^{88} &=& \frac{131072}{315} \sqrt{\frac{2 \pi }{17017}} 
\frac{G \nu m}{c^2 R} e^{-8 i \phi } x^4 \left\{ 1 - 7 \nu + 14 \nu^2 - 7 \nu^3
- x \left( \frac{3653}{342} - \frac{9325}{114} \nu + \frac{22351}{114} \nu^2 
- \frac{9107}{57} \nu^3 + \frac{4081}{114} \nu^4 \right) 
\right. \nonumber \\* && \left. + O(\epsilon^{3/2}) 
\right\}, \\
h^{87} &=& \frac{117649 i}{3240} \sqrt{\frac{7 \pi}{4862}} 
\frac{G \nu \delta m}{c^2 R} 
e^{-7 i \phi } x^{9/2} \left\{ 1 - 6 \nu + 10 \nu^2 - 4 \nu^3
  + O(\epsilon) \right\}, \\
h^{86} &=& - \frac{1944}{35} \sqrt{\frac{3 \pi }{85085}} 
\frac{G \nu m}{c^2 R} e^{-6 i \phi } x^4 \left\{ 1 - 7 \nu + 14 \nu^2 - 7 \nu^3
- x \left( \frac{353}{38} - \frac{7897}{114} \nu + \frac{18067}{114} \nu^2 
- \frac{6727}{57} \nu^3 + \frac{2653}{114} \nu^4 \right) 
\right. \nonumber \\* && \left. + O(\epsilon^{3/2}) 
\right\},
\end{eqnarray}
\begin{eqnarray}
h^{85} &=& - \frac{15625 i}{4536} \sqrt{\frac{5 \pi}{4862}} 
\frac{G \nu \delta m}{c^2 R} 
e^{-5 i \phi } x^{9/2} \left\{ 1 - 6 \nu + 10 \nu^2 - 4 \nu^3
  + O(\epsilon) \right\}, \\
h^{84} &=& \frac{1024}{4095} \sqrt{\frac{2 \pi }{935}} 
\frac{G \nu m}{c^2 R} e^{-4 i \phi } x^4 \left\{ 1 - 7 \nu + 14 \nu^2 - 7 \nu^3
- x \left( \frac{2837}{342} - \frac{6877}{114} \nu + \frac{15007}{114} \nu^2 
- \frac{5027}{57} \nu^3 + \frac{1633}{114} \nu^4 \right) 
\right. \nonumber \\* && \left. + O(\epsilon^{3/2}) 
\right\}, \\
h^{83} &=& \frac{81 i}{3640} \sqrt{\frac{3 \pi}{374}} 
\frac{G \nu \delta m}{c^2 R} 
e^{-3 i \phi } x^{9/2} \left\{ 1 - 6 \nu + 10 \nu^2 - 4 \nu^3
  + O(\epsilon) \right\}, \\
h^{82} &=& - \frac{8}{45045} \sqrt{\frac{\pi }{17}} 
\frac{G \nu m}{c^2 R} e^{-2 i \phi } x^4 \left\{ 1 - 7 \nu + 14 \nu^2 - 7 \nu^3
- x \left( \frac{2633}{342} - \frac{6265}{114} \nu + \frac{13171}{114} \nu^2 
- \frac{4007}{57} \nu^3 + \frac{1021}{114} \nu^4 \right) 
\right. \nonumber \\* && \left. + O(\epsilon^{3/2}) 
\right\}, \\
h^{81} &=& - \frac{i}{92664} \sqrt{\frac{\pi}{1190}} 
\frac{G \nu \delta m}{c^2 R} 
e^{-i \phi } x^{9/2} \left\{ 1 - 6 \nu + 10 \nu^2 - 4 \nu^3
  + O(\epsilon) \right\}.
\end{eqnarray}
\end{widetext}

In Appendix~\ref{sec:1pn-spin-weighted}, spin-weighted harmonic
components for even (odd) $\ell + m$ are given to 1PN (0PN) order for
arbitrary $\ell$.  Again, the $m<0$ components are given
by~(\ref{eq:mNegative}).


\subsection{Absorbing amplitude terms into the phase}
\label{sec:absorb-ampl-terms}

The $\ln{(x/x_0)}$ terms that appear in the spin-weighted spherical
harmonic components can be absorbed into a redefinition of the phase
by introducing an auxiliary phase variable $\psi = \phi + \delta$.
Since the $\ln{(x/x_0)}$ terms first enter at 1.5PN order, it is
straightforward to show that choosing~\cite{Blanchet96,Arun2004}
\begin{equation}
\label{eq:PhaseShift}
\delta =  - 3 \frac{M}{m} x^{3/2} \ln{\left(\frac{x}{x_0}\right)},
\end{equation}
where $M = {\cal I}$ (the mass monopole of the source), will
eliminate the $\ln{(x/x_0)}$ terms from the components.  This follows from 
\begin{eqnarray*}
h^{\ell m} &=& {\tilde h}^{\ell m} e^{-i m \psi} \\
&=& {\tilde h}^{\ell m} e^{-i m \phi} e^{-i m \delta} \\
&=& {\tilde h}^{\ell m} e^{-i m \phi} [ 1 - i m \delta 
- \frac{1}{2}m^2 \delta^2 + O(x^{9/2}) ],
\end{eqnarray*}
where ${\tilde h}^{\ell m}$ is $h^{\ell m}$ omitting the $\ln{(x/x_0)}$ terms.
Furthermore, since the orbital phase as a function of frequency goes
as 
\begin{equation}
\phi = - \frac{1}{32 \nu} x^{-5/2} + O(\epsilon).
\end{equation}
at leading order, the $\ln{(x/x_0)}$ terms, which were 1.5PN, 2.5PN,
and 3PN order in the amplitude terms, now appear as phase corrections
at relative order 4PN, 5PN, and 5.5PN.  As these terms are beyond the
order to which the orbital phase evolution is known (3.5PN order), it
can be argued that these terms can be ignored. Note that the choices
of $x_0$ in Eq.~(\ref{eq:LogParameter}) and $\delta$ in
Eq.~(\ref{eq:PhaseShift}) are not unique as other amplitude terms can
be absorbed into the phase (e.g. see~\cite{Kidder07}); these choices
were made to gather all logarithmic terms into one term, as well as to
simplify the waveform~\cite{Blanchet96}.

In order to recover the 2.5PN polarization waveforms
in~\cite{Arun2004,Kidder07} from the components listed above, it is
necessary to substitute $\ln{(x/x_0)}\to0$.  After substituting the
coefficients into Eq.~(\ref{eq:Decomposition}) and truncating the sum
at 2.5PN order, the result must be evaluated at $\Theta = i$, $\Phi =
\pi/2$.  Furthermore, there is an overall sign difference due to a
different choice of the polarization triad $({\vec N},{\vec P},{\vec
Q})$.

The polarization waveforms have been computed in the limit $\nu \to 0$
in~\cite{Tagoshi94} using black hole perturbation theory.  In order to
compare with their results, it is necessary to substitute 
\[ \ln{\left(\frac{x}{x_0}\right)} \to -\frac{17}{18} + \frac{2}{3} \ln{2}, \]
into the $h^{\ell m}$ listed above as~\cite{Tagoshi94} makes a
different choice in redefining the phase variable (and works in
Schwarzschild coordinates as opposed to harmonic coordinates).  After
this substitution, and setting $\nu=0$ and $\delta m/m = -1$, it is
found that the $h^{\ell m}$ above agree with the results of
~\cite{Poisson:1993vp,Tagoshi94}.\footnote{There is a sign difference
in $\zeta_+^{\ell,m}$ between~\cite{Poisson:1993vp}
and~\cite{Tagoshi94}.  The results presented here agree with the sign
of~\cite{Poisson:1993vp}.}


\section{Conclusions}
\label{sec:conclusions}

It has been shown that the spin-weighted spherical harmonic components
of the waveform can be computed to higher post-Newtonian order by
computing them directly from the radiative multipole moments rather
than by projecting them from the full polarization waveforms.  In
particular, this allows the dominant $h^{22}$ component to be computed
to 3PN order.  Since numerical simulations can compute the waveform
over the entire sphere, it is possible to compare the spin-weighted
spherical harmonic components of the waveform from the simulation with
those predicted by a quasi-adiabatic post-Newtonian inspiral.  Thus,
by examining the $h^{22}$ component, it can be determined whether or
not the 3PN contribution improves the agreement between the
post-Newtonian waveform and the numerical waveform.  If significant
improvement is found, it would suggest that it would be worth the
effort of computing the full 3PN waveform in order to improve
detection of marginal signals~\cite{VanDenBroeck06, VanDenBroeck07a,
Arun07a} as well as improve parameter extraction~\cite{Sintes00a,
Sintes00b, Moore02, Hellings03, VanDenBroeck07b, Arun07b, Trias07}.

In~\cite{Boyle2007} a high-accuracy comparison is made between
post-Newtonian generated waveforms and waveforms from a numerical
simulation of 15 orbits of an inspiral of an equal-mass non-spinning
binary black hole system.  For this case,~\cite{Boyle2007} finds that
the 3PN contributions to the amplitude of the $h^{22}$ mode improve
the accuracy with respect to the numerical waveforms.  This suggests
that for accurate parameter estimation, it may be desirable to compute
the full 3PN amplitude for the polarization waveforms.

For an equal-mass, non-spinning binary, only the source current
octupole needs to be computed to have the full polarization waveform.
For a non-spinning binary with an arbitrary mass ratio, much more
effort is required as explained at the end of~\cite{Arun2004}.  But
for comparison with numerical simulations, post-Newtonian theorists
should keep in mind that extending the PN order of a given radiative
multipole moment, will result in corresponding improvements in the
spin-weighted spherical harmonic components of the waveform, and these
corrections will be of interest even if all the corrections needed to
improve the polarization waveforms to the next order have yet to be
computed.


\begin{acknowledgments}
I would like to thank Michael Boyle, Gregory Cook, Abdul Mrou\'{e} and
Saul Teukolsky for helpful discussions concerning this work.  This
work was supported in part by a grant from the Sherman Fairchild
Foundation, by NSF grants PHY-0652952, DMS-0553677, PHY-0652929, and
NASA grant NNG05GG51G.
\end{acknowledgments}


\appendix


\section{1PN Source Mass Moments}
\label{sec:1pn-source-mass}

The source mass multipole moments for a system of $N$ (nonrotating)
compact point-masses to 1PN order is given
by~\cite{Blanchet89,Blanchet95a,Blanchet02}
\begin{eqnarray}
{\cal I}_L &=& \sum_{A=1}^N \left\{ {\tilde \mu_A} y_{A}^{<L>} +
\frac{1}{2 (2 \ell + 3) c^2} \frac{d^2}{dt^2} \left[ m_A y_A^2 
y_{A}^{<L>} \right] \right. \nonumber \\* && \left. 
- \frac{4 (2 \ell + 1)}{(\ell + 1) (2 \ell + 3) c^2}
\frac{d}{dt} \left[ m_A v_A^a y_A^{<aL>} \right] + O(\epsilon^2) \right\},
\nonumber \\ &&
\end{eqnarray} 
where $m_A$, ${\vec y_A}$, and ${\vec v_A}$ are the mass, position, and
velocity respectively of the $A$-th point mass, and
\begin{equation}
{\tilde \mu_A} = m_A \left[ 1 + \frac{3}{2} \frac{v_A^2}{c^2} -
\frac{G}{c^2} \sum_{B \neq A} \frac{m_B}{|{\vec y_A}-{\vec y_B}|}
 + O(\epsilon^2) \right].
\end{equation}
Restricting to the case of two bodies in a quasi-circular orbit, and
transforming to the center-of-mass frame using
\begin{eqnarray}
{\vec y_1} &=& \frac{m_2}{m} {\vec x} \left[ 1 + O(\epsilon^2) \right], \\
{\vec y_2} &=& - \frac{m_1}{m} {\vec x} \left[ 1 + O(\epsilon^2) \right],
\end{eqnarray}
and using the relative equation of motion~(\ref{eq:EOMa}) to eliminate
the relative acceleration from the derivatives yields
\begin{widetext}
\begin{equation}
\label{eq:SourceMassMoments1PN}
{\cal I}_L = 
\nu {\tilde m} \left\{ \left[ f_{\ell - 1}(\nu) - \gamma f_{\ell}(\nu) +
\gamma \frac{5 \ell^2 + 6 \ell + 9}{2 (\ell + 1) (2 \ell +3)} f_{\ell + 1}(\nu)
\right] x_{<L>} + \frac{\ell (\ell - 1) (\ell + 9)}{2 (\ell + 1) (2 \ell +3)} 
f_{\ell + 1}(\nu) \frac{r^2}{c^2} x_{<L-2} v_{i_{\ell-1} i_\ell >} 
+ O(\epsilon^2) \right\} 
\end{equation}
\end{widetext}
where 
\begin{equation}
\label{eq:OddEven}
\{ {\tilde m}, f_k (\nu) \} =
\begin{cases}
\{ m, s_k(\nu)\} & \text{for $\ell$ even} \\
\{ - \delta m, d_k(\nu)\} & \text{for $\ell$ odd} \\
\end{cases}
\end{equation}
where $s_\ell = (m_1^\ell + m_2^\ell)/m^\ell$ and $d_\ell = (m_1^\ell
- m_2^\ell)/m^\ell$, which themselves can be rewritten as polynomials
in $\nu$ as
\begin{eqnarray}
s_\ell(\nu) &=& 1 + \sum_{k=1}^{\ell/2}
\left[ \binom{\ell - k - 1}{k - 1} + \binom{\ell - k}{k} \right] 
 (-\nu)^k, \nonumber \\* && \\
d_\ell(\nu) &=& \sum_{k=0}^{\ell/2} \binom{\ell - k - 1}{k} (-\nu)^k .
\end{eqnarray}


\section{0PN Source Current Moments}
\label{sec:0pn-source-current}

The source current multipole moments for a system of N (nonspinning)
compact point-masses is given to 0PN order by~\cite{Damour:1990ji,Blanchet02}
\begin{equation}
{\cal J}_L = \sum_{A=1}^N m_A \epsilon^{ab<i_\ell} y_A^{L-1> a} v_A^b 
+ O(\epsilon).
\end{equation}
Restricting to the case of two bodies in a quasi-circular orbit, and
transforming to the center-of-mass frame as for the mass moments in
Appendix~\ref{sec:1pn-source-mass}, it is straightforward to show that
the source current multipole moments to leading order are given by
\begin{equation}
\label{eq:SourceCurrentMoments0PN}
{\cal J}_L = \frac{\hat m}{m} g_\ell(\nu) L_{<i_\ell} x_{L-1>} + O(\epsilon), 
\end{equation}
where 
\begin{equation}
\label{eq:OddEvenJ}
\{ {\hat m}, g_k (\nu) \} =
\begin{cases}
\{ - \delta m, d_k(\nu)\} & \text{for $\ell$ even} \\
\{ m, s_k(\nu)\} & \text{for $\ell$ odd}
\end{cases}
\end{equation}


\section{1PN Spin-Weighted Spherical Harmonic Components}
\label{sec:1pn-spin-weighted}

\subsection{$\ell + m$ even}

For even $\ell + m$ (for non-spinning binaries), the spin-weighted
spherical harmonic components are due to contributions from the
radiative mass multipoles.  To 1PN order, the radiative moments are
given simply $\ell$ time derivatives of the source moments.  For
arbitrary $\ell$, the mass moment is given by
Eq.~(\ref{eq:SourceMassMoments1PN}).  In evaluating the spin-weighted
spherical harmonic components from the source mass moments, the following
identity is useful:
\begin{eqnarray}
x_L {\cal Y}_L^{\ell m} &=& r^\ell n_L {\cal Y}_L^{\ell m} \nonumber \\ 
&=& r^\ell Y^{\ell m}\left(\frac{\pi}{2},\phi\right),
\end{eqnarray}
where $\phi$ is the orbital phase (as opposed to an angle on the sphere).
Since $\phi = \omega t$, 
\begin{equation}
\frac{d^k}{dt^k}  Y^{\ell m}\left(\frac{\pi}{2},\phi\right) =
(-i m \omega)^k  Y^{\ell m}\left(\frac{\pi}{2},\phi\right).
\end{equation}
Therefore, it can be shown that 
\begin{equation}
x_{L-2} v_{i_{\ell-1} i_\ell}  {\cal Y}_L^{\ell m} = 
\frac{\ell - m^2}{\ell (\ell-1)} \omega^2 
r^\ell Y^{\ell m}\left(\frac{\pi}{2},\phi\right),
\end{equation}
using that ${\dot r}$ and ${\dot \omega}$ are of 2.5 order.
Finally, using that
\begin{equation}
\left( \frac{r \omega}{c} \right)^\ell = x^{\ell/2} \left[ 1 - x \ell \left(
1 - \frac{1}{3} \nu \right) \right],
\end{equation}
it can be shown that
\begin{widetext}
\begin{eqnarray}
h^{\ell m} &=& (i m)^\ell 
\frac{8 \pi}{(2 \ell + 1)!!} \sqrt{\frac{(\ell + 1)(\ell + 2)}
{\ell (\ell - 1)}} \frac{G \nu {\tilde m}}{c^2 R} x^{\ell/2}
Y^{\ell, -m} \left(\frac{\pi}{2},\phi\right) \left\{ f_{\ell - 1}(\nu) \left[
1 - x \ell \left( 1 - \frac{1}{3} \nu \right) \right] + 
\frac{3}{2} x f_{\ell+1}(\nu) \right. \nonumber \\* && \left. - x f_\ell(\nu) -
x \frac{m^2 (\ell + 9)}{2 (\ell + 1) (2 \ell +3)} f_{\ell+1}(\nu) 
+ O(\epsilon^{3/2}) \right\}
\qquad \text{(for $\ell + m$ even)},
\end{eqnarray}
\end{widetext}
where ${\tilde m}$ and $f_k(\nu)$ are given by~(\ref{eq:OddEven}).

\subsection{$\ell + m$ odd}

For odd $\ell + m$ (for nonspinning binaries), the spin-weighted
spherical harmonic components are due to contributions from the
radiative current multipoles.  To 0PN order, the radiative moments are
given simply $\ell$ time derivatives of the source moments.  For
arbitrary $\ell$, the current moment is given by
Eq.~(\ref{eq:SourceCurrentMoments0PN}).  In evaluating the
spin-weighted spherical harmonic components from the source current
moments, the following identity is useful:
\begin{equation}
\delta_{z i_\ell} n_{L-1}  {\cal Y}_L^{\ell m} =
 \frac{1}{\ell} \sqrt{\frac{(2 \ell + 1)(\ell^2 - m^2)}{2 \ell - 1}} 
Y^{\ell-1, m}\left(\frac{\pi}{2},\phi\right).
\end{equation}

Following the same steps as above, it can be shown that
\begin{widetext}
\begin{equation}
h^{\ell m} = - (i m)^\ell 
\frac{16 \pi i}{(2 \ell + 1)!!} \sqrt{\frac{(2 \ell + 1)(\ell + 2)
(\ell^2 - m^2)}{\ell (2 \ell - 1)(\ell + 1) (\ell - 1)}} 
\frac{G \nu {\hat m}}{c^2 R} x^{(\ell+1)/2}
Y^{\ell-1, -m} \left(\frac{\pi}{2},\phi\right) \left[ g_\ell(\nu) + O(\epsilon)
\right] \qquad \text{(for $\ell + m$ odd)},
\end{equation}
\end{widetext}
where ${\hat m}$ and $g_k(\nu)$ are given by~(\ref{eq:OddEvenJ}).





\bibliography{References/References}


\end{document}